\documentclass[envcountsect]{llncs}
\usepackage{hyperref}
\usepackage{amsmath}
\usepackage{graphicx}
\usepackage{caption}
\usepackage{amssymb}
\usepackage{subfig}
\usepackage{verbatim}
\usepackage{array}
\usepackage{multirow}
\usepackage{color}
\usepackage{color,soul}
\usepackage{float}
\usepackage{makecell}
\usepackage{multirow}

\begin{document}

\title{Privacy-Preserving Link Prediction}

\author{Didem Demirag\inst{1} \and Mina Namazi \inst{2} \and  Erman Ayday\inst{3} \and Jeremy Clark \inst{1}}
\institute{	Concordia University \\Montreal, QC, Canada \\ \email{	d\_demira@encs.concordia.ca, j.clark@concordia.ca} \and Open University of Catalonia\\ Barcelona, Spain \\ \email {mnamaziesfanjani@uoc.edu} \and 	Case Western Reserve University, \\Cleveland, OH, USA \\ \email{	exa208@case.edu}  }

\maketitle

\begin{abstract}

Consider two data holders, ABC and XYZ, with graph data (\textit{e.g.,} social networks, e-commerce, telecommunication, and bio-informatics). ABC can see that node A is linked to node B, and XYZ can see node B is linked to node C. Node B is the common neighbour of A and C but neither network can discover this fact on their own. In this paper, we provide a two party computation that ABC and XYZ can run to discover the common neighbours in the union of their graph data, however neither party has to reveal their plaintext graph to the other. Based on private set intersection, we implement our solution, provide measurements, and quantify partial leaks of privacy. We also propose a heavyweight solution that leaks zero information based on additively homomorphic encryption. 

\end{abstract}


\newcommand{\textblue}[1]{\textcolor{blue}{#1}}
\newcommand{\textred}[1]{\textcolor{red}{#1}}

\section{Introduction}\label{sec:Introduction}

Link prediction discovers important linkages between nodes in a graph. Based on the analysis of these linkages, it helps the data holder to forecast what future connections might emerge between the nodes, and to predict if there are missing links in the data. Some common applications include: (i) in social networks, to recommend links between users; (ii) in e-commerce or personalized advertisement, to recommend products to users; (iii) in telecommunication, to build optimal phone usage plans between the users; and (iv) in bioinformatics, to predict associations between diseases and attributes of patients or to discover associations between genes (or proteins) and different functions. 

Link prediction is typically done on the local graph of a data holder or service provider. For instance, a social network, analyzing the common neighbours between its users decides whether to recommend links between the users. However, link prediction will be more accurate and correct by considering more information about the graph nodes. This can be achieved by merging two or more graph databases that include similar information, leading to ``distributed link prediction'' between two or more graph databases. For instance, two social networks may utilize the connections in their combined graph to provide more accurate link prediction for their users. Furthermore, distributed link prediction will enable different uses of link prediction, such as building connections between users and products based on the tastes of other similar users (\textit{e.g.,} friends of a user). Such an application may be possible between graph databases of a social network and an e-commerce service provider. In some cases, collaboration is mutually beneficial to both parties. In others, one party can pay the other party to participate---one party gets better data and the other gets to monetize its data.

Distributed link prediction, although is a promising approach for more accurate and richer link prediction applications, also results in privacy concerns since it implies combining two or more different graph databases. In this scenario, threats against privacy can be categorized into three groups~\cite{surveyofprivacy}: identity disclosure, link disclosure, and attribute disclosure. All these threats should be considered in a distributed link prediction algorithm, since it involves privacy-sensitive databases from multiple parties. 

One promising solution for this privacy concern is cryptography to achieve distributed link prediction in a privacy-preserving way. Thus, in this paper, our goal is to develop a cryptographic solution for privacy-preserving distributed link prediction between multiple graph databases. We propose a solution based on private set intersection (PSI) to tackle this problem by considering both the efficiency of the solution and its privacy. Via evaluations, we show that this solution provides good efficiency. For example, it can run in under 1 second (ignoring communication latency) for graphs based on a Flickr dataset with ~40K nodes. The proposed protocol does not provide perfect privacy (it leaks some intermediary values) and so we quantify this leakage to better understand if it is consequential enough to move to a fully private solution (which we also sketch).

\subsection{Use Cases}
Privacy-preserving distributed link prediction can be utilized in different settings. Here, we explain some of the possible applications.

\paragraph{Social networks.} In this setting, there are two social networks, Graph 1 and Graph 2. Graph 1 aims to understand whether there will be a link formed between nodes $x$ and $y$ by also utilizing the similarity of $x$ and $y$ in Graph 2, as distributed link prediction provides better accuracy compared to performing this operation locally.

\paragraph{E-commerce.} Another application can be between a social network and an e-commerce service. In the previous use case, the link between two users is the main concern of the protocol. Unlike the previous case, here the links between a user and products are determined at the end of the protocol. 
In the e-commerce graph, there are links between the users and the products that they have bought. The aim here is to provide better advertising to users. The network will recommend product $n$ to the user $x$ if this user's friends also purchased the same product. For this purpose, the e-commerce network has to know the friends of user $x$ in the social network. Unlike the previous use case, here link prediction cannot be done locally on the e-commerce graph, as the knowledge of the social network's structure should be utilized in order to do the recommendation.

\paragraph{Telecommunication.} 
In this use case, an advertising company wants to propagate an advertisement in the telecom network. If user $x$ is a target for that advertisement, the company would like to know which nodes are likely to form links with user $x$, in order to decide which nodes it will send the advertisement. The aim is to maximize the number of nodes that learn about the advertisement. Another application involves a social network graph and a phone operator graph. The phone operator wants to find out friends of user $x$ in the social network, so that it offers the special services (e.g., discounts) to the users that are similar to user $x$. 

\paragraph{Bioinformatics.} Here, the first graph consists of patients and diseases and the aim is to predict the link between the patient $i$ and the disease $j$. In the second graph, there are similar patients to patient $i$. Using these similar patients, and their connection to disease $j$, the link between patient $i$ and disease $j$ can be inferred. 

\subsection{Related Work}\label{sec:RelatedWork}

 There is a rich literature on link prediction algorithms (without consideration of privacy) in a variety of network structures: multiple partially aligned social networks~\cite{zhang2014meta}; coupled networks~\cite{dong2015coupledlp}; and heterogenous networks ~\cite{tangtransfer}. Other works consider node similarity when two nodes in the graph do not share common neighbours~\cite{leicht2006vertex}; unbalanced, sparse data across multiple heterogeneous networks~\cite{dong2012link}; missing link prediction using local random walk~\cite{liu2010link}; and the intersection of link prediction and transfer learning~\cite{yu2008gaussian}. 

Other research considers the use of cryptography for collaborating on graph-based data between two parties with privacy protections. However such works consider problems other than link prediction: merging and query performed on knowledge graphs owned by different parties~\cite{chen2020survey}; whether one graph is a subgraph of the other graph~\cite{xu2019privacy}; single-source shortest distance and all-pairs shortest distance both in sparse and dense graphs~\cite{anagreh2021parallel}; all pairs shortest distance and single source shortest distance~\cite{privpresgraph}; and transitive closure~\cite{he2009efficient}; anonymous invitation-based system~\cite{boshrooyeh2017inonymous} and its extension to malicious adversarial model~\cite{boshrooyeh2019anonyma}. While it may be possible to transform some of these into finding common neighbours with a black-box approach, we provide a purpose-built protocol for common neighbour. Later in Section~\ref{sec:building}, we review potential cryptographic building blocks in the literature.

\section{Proposed Solution}\label{sec:method}

\subsection{Building Blocks from Data Mining}\label{sec:Backg}

\paragraph{Link prediction.}
Given a snapshot of a graph at time $t$, link prediction algorithms aim to accurately predict the edges that will be added to the graph during the interval from time $t$ to a given future time $t'$~\cite{linkepred}.

\begin{table}[t]
	\begin{center} 
		\scalebox{0.9}{
			\begin{tabular}{l*{6}{c}r}
				\textbf{Similarity metric }& \textbf{Definition}\\
				\hline
				Common neighbours & $|\Gamma(x)\cap \Gamma(y)|$ \\
				\hline
				Jaccard's coefficient           & $\frac{|\Gamma(x)\cap \Gamma(y)|}{|\Gamma(x)\cup \Gamma(y)|} $\\
				\hline
				Adamic/Adar & $\sum_{z\epsilon |\Gamma(x)\cap \Gamma(y)|}{\frac{1}{log(|\Gamma(z)|)}}$\\
				\hline
				\multirow{4}{*}{$Katz_\beta$ } &  $\sum_{l =1 }^{\infty}{\beta^l . |path_{x,y}^{\langle l \rangle}|}$\\
				&where $path_{x,y}^{\langle l \rangle}:=\{$ paths of length exactly $l$ from $x$ to $y$\} \\
				&weighted: $ path_{x,y}^{\langle l \rangle} :=$ weight of the edge between $x$ and $y$\\
				&unweighted: $ path_{x,y}^{\langle l \rangle} := 1$ iff $x$ and $y$ are 1-hop neighbours\\
				&The weight is determined by the constant value $\beta$. \\\hline
				
		\end{tabular}}
	\end{center}
	\caption {Different similarity metrics in a graph.\label{Table:sim}}
\end{table}

\paragraph{Similarity metrics.}
In Table~\ref{Table:sim}, different metrics for calculating proximity are given. Common neighbours, Jaccard coefficient and Adamic-Adar index are regarded as the node-dependent indices and they only require the information about node degree and the nearest neighbourhood, whereas the Katz index is defined as path-dependent index that consider the global knowledge of the network topology~\cite{liu2010link}. While there are also other metrics that are used (some of which are shown in Table~\ref{Table:sim}), we choose common neighbours, as it is one of the widely-used methods for link prediction.

\paragraph{Common Neighbours.}

Common neighbours is used to predict the existence of a link between two nodes based on the number of their common neighbours. If two nodes share common neighbours, it is more likely that they will be connected in the future. In a local graph, the result of the metric can directly be computed by determining the intersection of the neighbour sets of two nodes. Based on the cardinality of the set, the network decides whether to suggest a link between these two nodes. The cardinality is defined as
\[ \mathrm{common~neighbours} = |\Gamma(x)\cap \Gamma(y)| \], 
where $ \Gamma(x) $ and $ \Gamma(y) $ are the set of neighbours of nodes $x$ and $y$ respectively.

\paragraph{Adapting Common Neighbours for Two Parties.}

For the use case in this paper, we look at the problem of computing common neighbours metric across two different graphs owned by different entities. For example, this could be two separate social networks, or a social network with an e-commerce network. Graph 1 wants to perform link prediction between the nodes $x$ and $y$ by using the common neighbours information from Graph 2. $\mathrm{CN}$ denotes the total number of common neighbours that will be determined at the end of the protocol. We propose the following computation for two graphs, taking care to not double count any common neighbours:

\[\mathrm{CN} = \mathrm{local1 + local2 + crossover1 + crossover2 - overlap} \]

The variables are as follows:

\begin{itemize}
    \item local1: number of common neighbours of node x and node y in Graph 1
    \item local2: number of common neighbours of node x and node y in Graph 2
    \item crossover1: number of common neighbours of node x from Graph 1 and node y from Graph 2
    \item crossover2: number of common neighbours of node y from Graph 1 and node x from Graph 2
    \item overlap: intersection of local1 and local2
\end{itemize}

\begin{figure}[t]
	\centering
	\includegraphics[width=.95\textwidth]{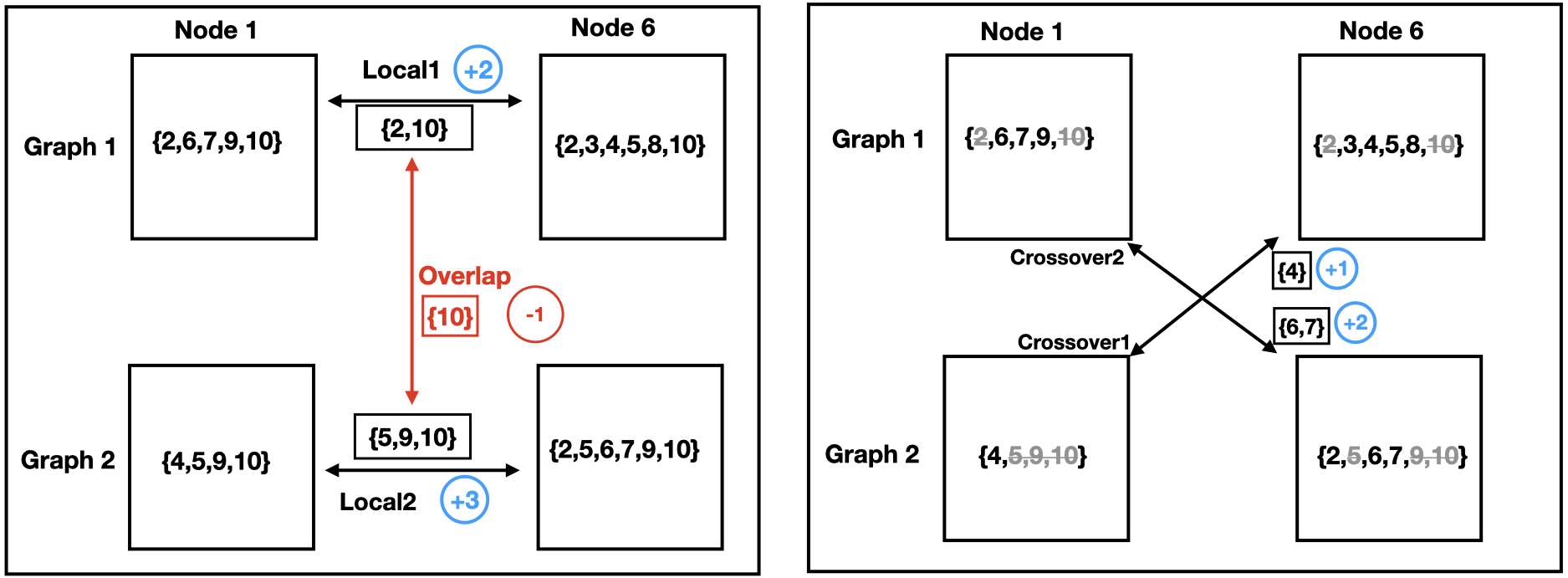}
	\caption{An example computation between Graph 1 and Graph 2 to find the number of common neighbours of nodes 1 and 6 in their joint graph. Our contribution is to perform this computation in a privacy-preserving manner.\label{fig:CNexample}}
\end{figure}

Figure~\ref{fig:CNexample} illustrates an example of how CN is computed using the neighbours sets of both Graph 1 and Graph 2 (based on the graphs in Figure~\ref{fig:info}). Graph 1 decides whether to suggest a link between nodes 1 and 6 based on this cardinality.

\begin{table}
	
	\begin{tabular}{|m{1em}|l|l|}
		\hline
		\multirow{25}{*}{\rotatebox{90}{PSI based}} &\multicolumn{1}{|p{0.3\textwidth}|}{ PSI~\cite{de2012fast}} & \multicolumn{1}{|p{0.7\textwidth}|}{\footnotesize \textbf{Complexity:} Protocol complexity is linear in the sizes of the two sets. Both the client and the server performs exponentiations and modular multiplications.
			\newline  \textbf{Info leaked: } Intersection cardinality, no third party \newline  \textbf{Security setting: }semi-honest  } \\ \cline{2-3} 
		& \multicolumn{1}{|p{0.3\textwidth}|}{Delegated PSI~\cite{duong2020catalic}  }  &  \multicolumn{1}{|p{0.7\textwidth}|}{\footnotesize \textbf{Complexity: } Computation and communication complexity of the protocol is linear in the size of the smaller set. For polynomial interpolation field operations are performed. Cloud server has to evaluate oblivious distributed key PRF
			instances, and unpack messages. The waiting time of packing messages by the backend server is the main computation cost.  \newline  \textbf{Info leaked: }  Intersection cardinality, uses third party \newline  \textbf{Security setting: } semi-honest  } \\ \cline{2-3} 
		&\multicolumn{1}{|p{0.3\textwidth}|}{PSI with FHE~\cite{chen2017fast}  }  & \multicolumn{1}{|p{0.7\textwidth}|}{\footnotesize \textbf{Complexity: } Communication overhead is logarithmic in the larger set size and linear in the smaller set size.  While FHE is asymptotically efficient, it isn't in practice.
			\newline  \textbf{Info leaked: }Input sizes and bit string length of the sets, no third party \newline  \textbf{Security setting: }Semi-honest  }  \\ \cline{2-3} 
		&\multicolumn{1}{|p{0.3\textwidth}|}{  PSI with OT~\cite{pinkas2019efficient}  } &\multicolumn{1}{|p{0.7\textwidth}|}{\footnotesize \textbf{Complexity: } The circuit-based PSI protocol has linear communication complexity. 
			\newline  \textbf{Info leaked: } No info leaked as the result of a function on the intersection cardinality is the output, no third party \newline  \textbf{Security setting: } semi-honest}   \\ \cline{2-3} 
		& \multicolumn{1}{|p{0.3\textwidth}|}{Labeled PSI with FHE in malicious setting~\cite{chen2018labeled} }   & \multicolumn{1}{|p{0.7\textwidth}|}{\footnotesize \textbf{Complexity: }Communication overhead is logarithmic in the larger set size and linear in the smaller set size. While FHE is asymptotically efficient, it isn't in practice.  \newline  \textbf{Info leaked: } No info is leaked, as the output is secret shared, no third party\newline  \textbf{Security setting: } Malicious}  \\ \hline
		\multirow{1}{*}{\rotatebox{90}{Non-PSI based}}          & \multicolumn{1}{|p{0.3\textwidth}|}{Privacy-preserving integer comparison~\cite{health} over each pair}    & \multicolumn{1}{|p{0.7\textwidth}|}{\footnotesize \textbf{Complexity: } Privacy-preserving integer comparison protocol is run between every pair of nodes in the adjacency matrix created using the neighbour list from both graphs. The comparison protocol performs encryption, partial decryption, modular exponentiation, and multiplications.  \newline  \textbf{Info leaked:}  No info is leaked, no third party \newline  \textbf{Security setting: } Malicious if ZKP added  }  \\ \hline 
	\end{tabular}
	
	\caption{PSI-based and Non-PSI based cryptographic building blocks \label{Table:crypt}}
	
\end{table}


\subsection{System Model}

In our setting, there are two parties: Graph 1 and Graph 2, each having a graph structured network. Graph 1 wants to determine whether to suggest a link between the nodes $x$ and $y$, not by only determining the common neighbours using its own graph, but also utilizing the graph structure of Graph 2. Graph 1 and Graph 2 compute common neighbours on their joint graphs without disclosing their respective graph structures. The result (number of common neighbours) is provided only to Graph 1, however the protocol can be run twice if Graph 2 also wants the result (otherwise, we assume Graph 1 is paying Graph 2 for this service). While creating our scheme, we make the following assumptions:

\begin{enumerate}
	\item The identifiers in both graphs for the same nodes match. The graphs should prepare for the computation by sharing a schema and agreeing on unique identifiers (\textit{e.g.,} an email address or phone number for human users).
	
	\item Both graphs know the identity of the nodes for which the computation is being performed. In other words, edges involving these nodes are hidden, as well as all other edges and nodes.
	
	\item If $x$ and $y$ are direct neighbours in Graph 1, Graph 1 has no need for the computation.
	
	\item If $x$ and $y$ are direct neighbours in Graph 2, Graph 2 will halt before doing the computation and inform Graph 1. In this case, Graph 1 discovers a hidden link between $x$ and $y$, which is stronger for prediction than the number of common neighbours.
	
\end{enumerate}

\paragraph{Threat Model.} 

The common public input to the computation will be the identifiers of two nodes known to Graph 1 and Graph 2. The private input of Graph 1 and Graph 2, respectively, is an assertion of their graph data. We assume Graph 1 and Graph 2 honestly input their correct data. This is a common assumption and resolving it involves having the data authenticated outside of the protocol, which is not a natural assumption for our use-cases. The second question is whether we can assume Graph 1 and Graph 2 follow the protocol correctly (semi-honest model) or exhibit arbitrary behaviour (malicious model). Given the strong assumption of data input, we find it natural to fit it to a semi-honest model of the protocol.

With these assumptions, we design the protocol so that Graph 2 learns nothing about Graph 1 other than the common input. On the other hand, Graph 1 learns the number of common neighbours on the joint set, which is the output of the multiparty computation (MPC). A fundamental limitation of MPC is that the output itself can leak information about the input. For example, if Graph 1 is malicious and is able to repeat this protocol many times with Graph 2, it can slowly reconstruct Graph 2's input by adaptively providing different inputs each time. For the purposes of this paper, we assume Graph 1 will not do this, because it is semi-honest, and further Graph 2 would not entertain so many executions of the protocol.

As an artifact of our protocol, Graph 1 also learns the intermediate values to compute the number of common neighbours: local1 + local2 + crossover1 + crossover2 - overlap. This extra information does allow a malicious Graph 1 to reconstruct Graph 2 with fewer queries, but in Section~\ref{sec:leak} we show that the impact of the leakage is immaterial. This can be prevented with heavier cryptography (Section~\ref{sec:phe}). In addition, Graph 2 can force Graph 1 to compute the wrong result only if it behaves maliciously. In conclusion, our threat model provides reasonable privacy protection while being lightweight enough to be practical, and we suggest it represents a useful compromise for many real-world applications.

\subsection{Building Blocks from Cryptography}\label{sec:crypto}
\label{sec:building}

Private set intersection (PSI) is a two-party cryptographic protocol that allows two entities, each with a set of data, to learn the intersection of their data without either learning any information about data that is outside the intersection~\cite{freedman2004efficient}. After a detailed investigation of PSI variants and other related primitives, we choose \cite{de2012fast} as the core scheme to deploy for our link prediction. Our scenario requires a scheme that calculates only the cardinality (sometimes called PSI-CA) of the intersection of the two sets in an efficient and scalable way with minimum information leakage with no third party's assistance. A security model for semi-honest adversaries is sufficient, and we leave additional (stronger) security guarantees for future work. A summary of relevant cryptographic primitives is given in Table~\ref{Table:crypt} and we provide more details of each primitive in  Appendix~\ref{app:crypto}. 

\subsection{Proposed Protocol} 

\begin{figure}[t]
	\centering
	\includegraphics[width=.99\textwidth]{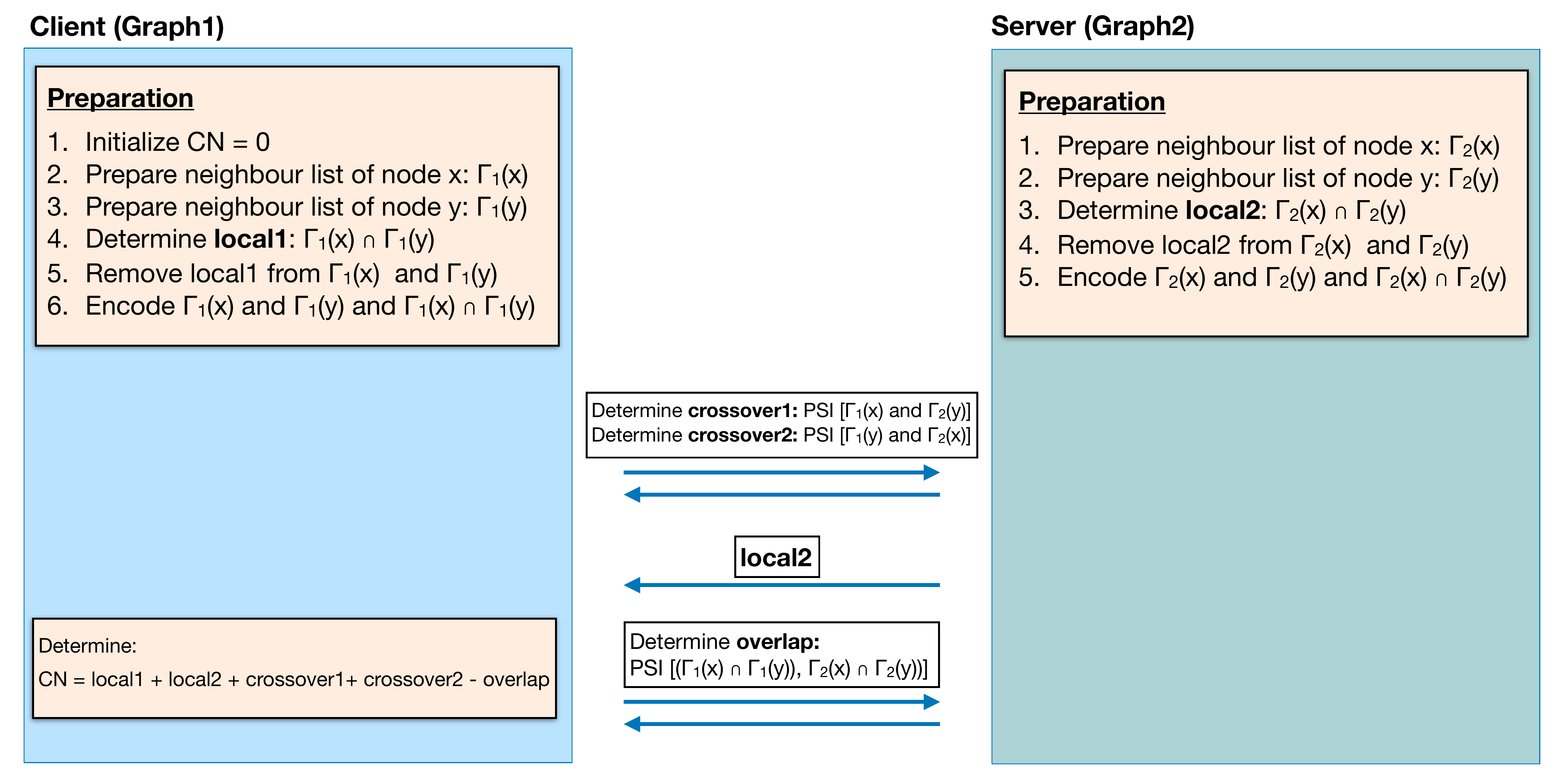}
	\caption{Overview of the proposed PSI-based solution. PSI is called three times in the protocol for determining crossover1, crossover2 and overlap. PSI itself is described in Figure~\ref{fig:PSIdetails}}
	\label{fig:CNv1}
	\vspace{-10pt}
\end{figure}

\begin{figure}[t]
	\centering
	\includegraphics[width=.99\textwidth]{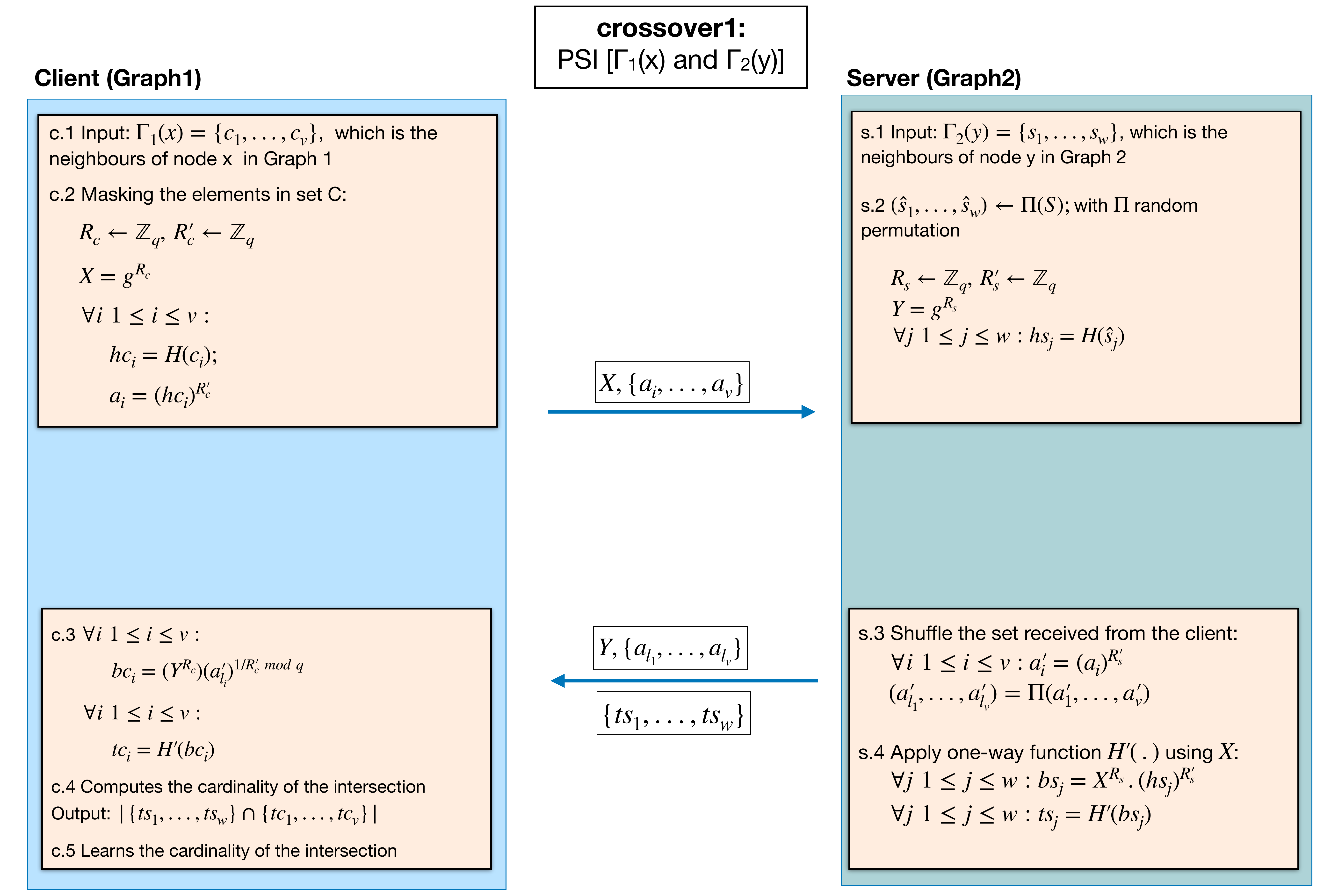}
	\caption{PSI protocol for determining crossover1 (adapted from~\cite{de2012fast})}
	\label{fig:PSIdetails}
	\vspace{-10pt}
\end{figure}

We use PSI scheme proposed in ~\cite{de2012fast} to perform distributed link prediction between two graph databases.  Figure~\ref{fig:CNv1} shows the interactive protocol between Graph 1 and Graph 2. Graph 1 wants to learn the common neighbour index to determine whether to suggest a link between the nodes $x$ and $y$. Both Graph 1 and Graph 2 locally determine the neighbour sets of $x$ and $y$ (local1 and local2, respectively). In order to determine crossover1, crossover2, and overlap, Graph 1 and Graph 2 run three separate PSI protocols among themselves. Each PSI leaks a certain amount of information and we discuss this partial information leak in Section~\ref{sec:leak}. At the end of the protocol, Graph 1 learns the exact cardinality of common neighbours of nodes $x$ and $y$ on the joint graph. Figure~\ref{fig:PSIdetails} shows the details of the PSI protocol for calculating crossover1 (the calculations for crossover2 and overlap are also the same). It is an interactive protocol between Graph 1 and Graph 2, with offline and online stages. At the offline stage, Graph 1 masks its set and Graph 2 masks its set and shuffles it. During the online stage, Graph 2 receives the masked set of Graph 1, masks it with its own randomness and shuffles it. When Graph 1 receives the sets, it removes the randomness and determines the intersection of two sets. PSI is described in the multiplicative subgroup $\mathbb{G}_q$ of $\mathbb{Z}_p^*$, where $p$ and $q$ are large primes, $q\mid p-1$ and $g ~ \epsilon ~\mathbb{G}_q$ is the generator. $|p|=1024$ bits and $|q|=160 $ bits. This is for experimental purposes only, these parameters should be at least doubled in length to meet the current, accepted security level~\footnote{\href{https://nvlpubs.nist.gov/nistpubs/SpecialPublications/NIST.SP.800-131Ar2.pdf}{NIST Special Publication 800-131A
Revision 2}}. H and H' are hash functions that are modeled as random oracles. 

\section{Evaluation}

\subsection{Performance}

\begin{figure}[t]%
  \captionsetup{justification=centering} 
    \subfloat[ Total run-time: Node 1's neighbours in Graph 1]{{\includegraphics[width=5cm]{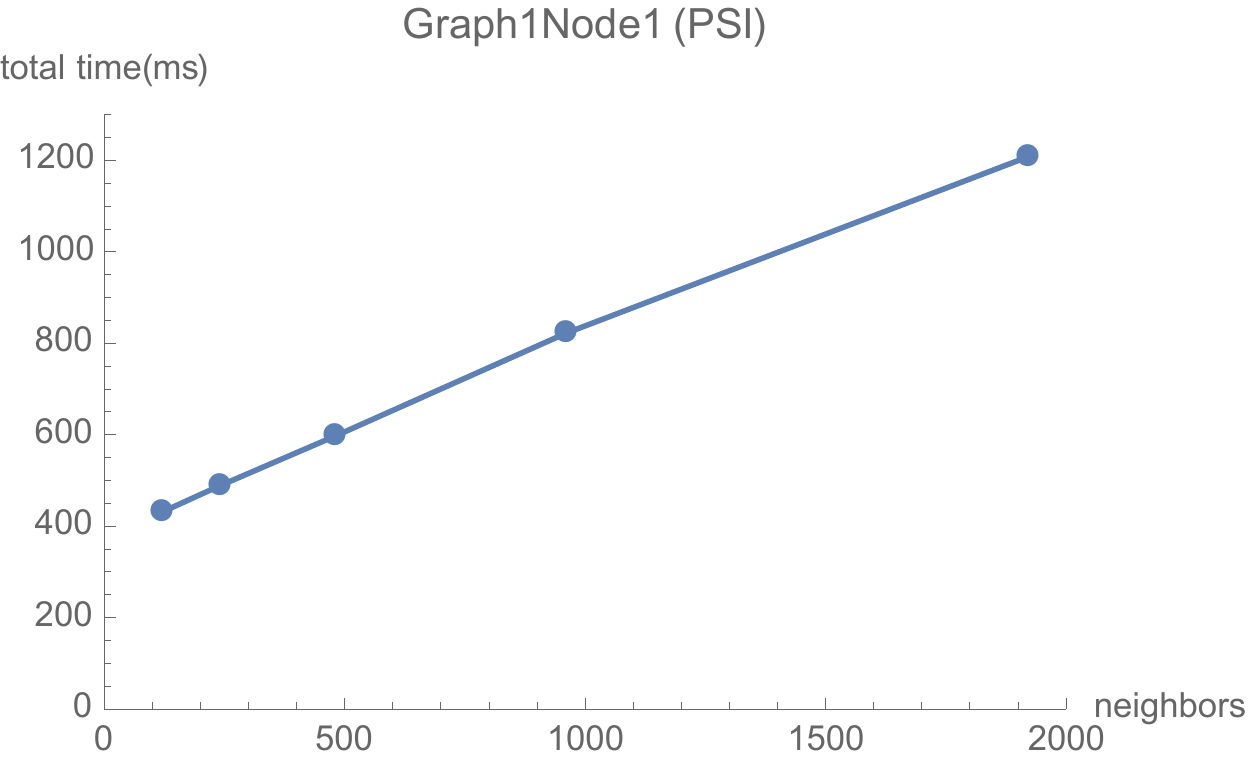} }}%
    \qquad
    \subfloat[ Total run-time: Node 2's neighbours in Graph 1 ]{{\includegraphics[width=5cm]{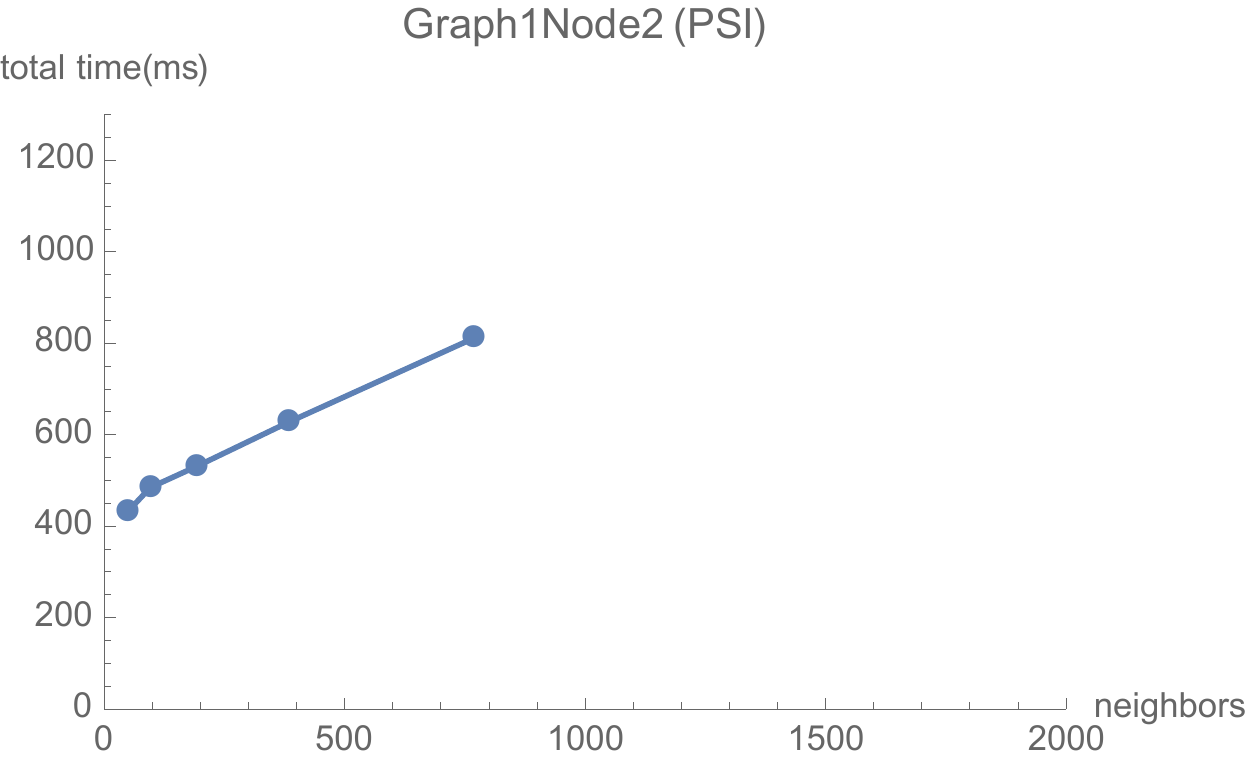} }}%
    \\
    \subfloat[ Total run-time: Node 1's neighbours in Graph 2]{{\includegraphics[width=5cm]{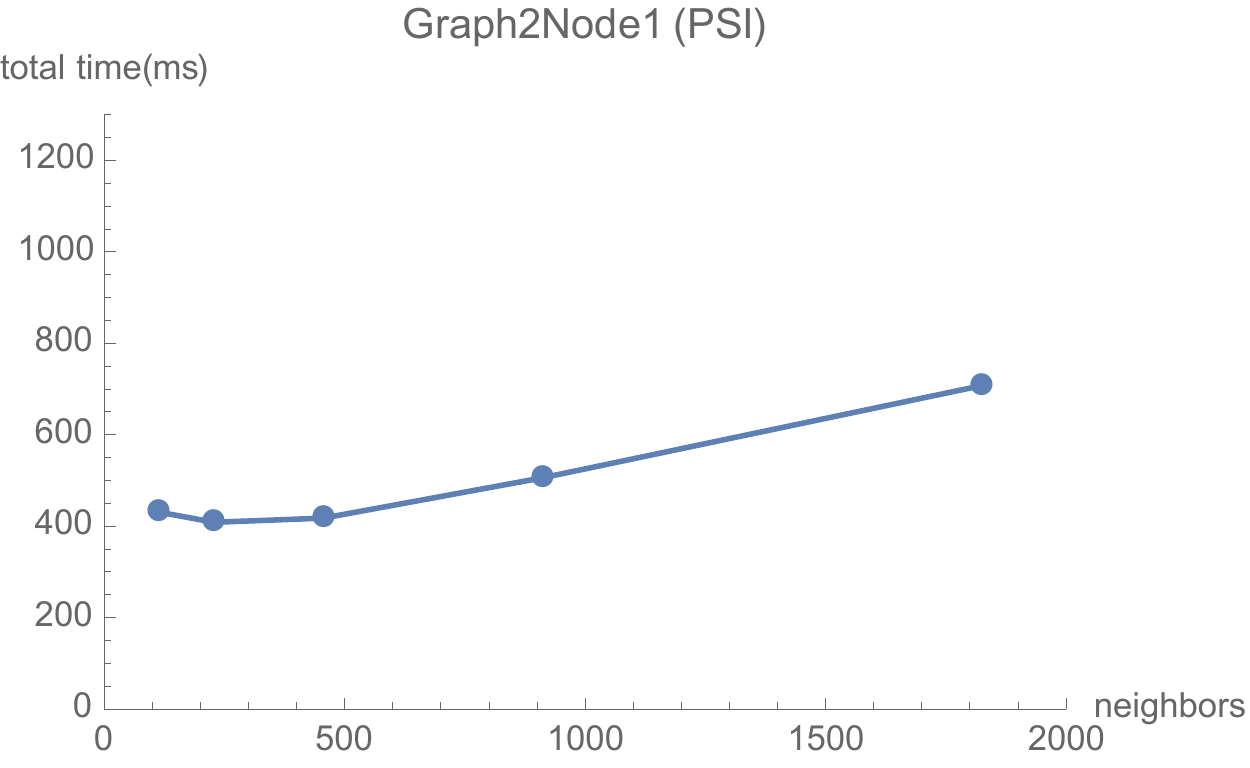} }}%
    \qquad
    \subfloat[ Total run-time: Node 2's neighbours in Graph 2]{{\includegraphics[width=5cm]{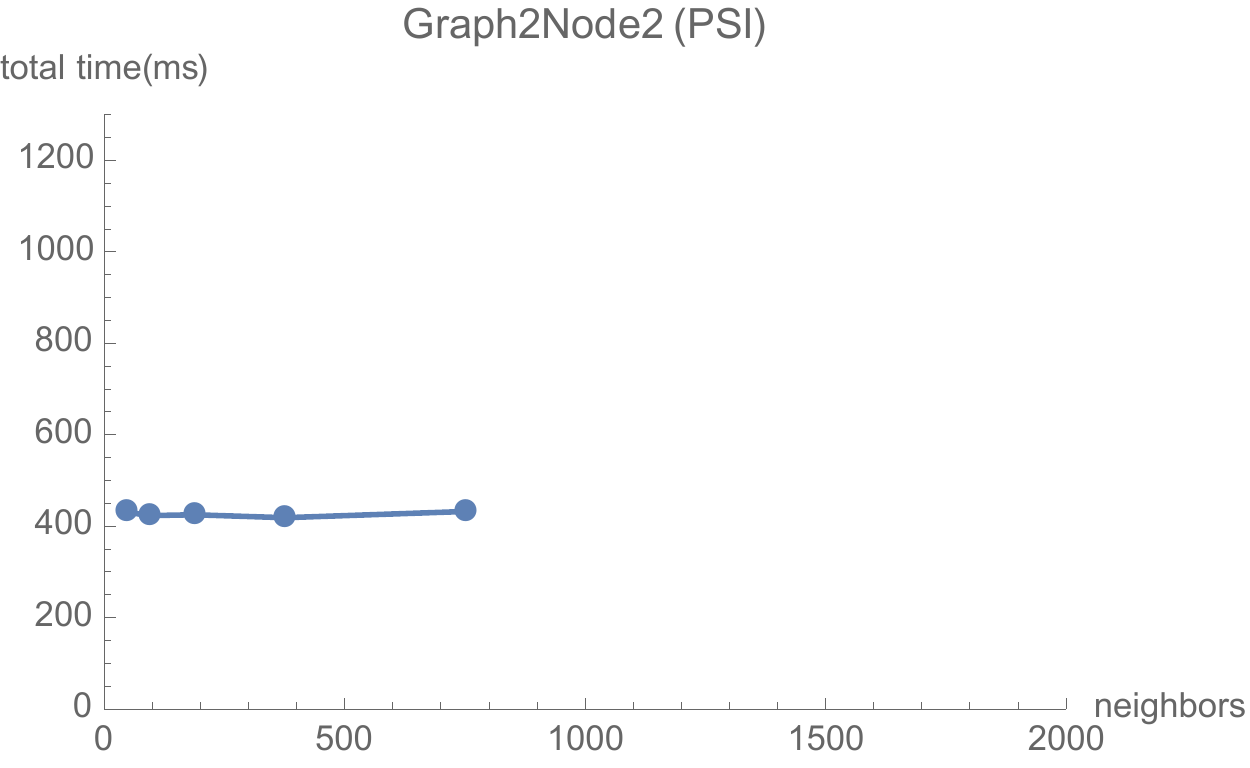} }}%
    \caption{Total run-time (in milliseconds) of common neighbour on joint graph according to varying sizes of neighbours for Node 1 and Node 2.\label{fig:experiments}}%
\end{figure}

We implemented the proposed distributed link prediction algorithm and evaluated it considering different aspects. We used the implementation of PSI defined in ~\cite{de2012fast} where $q$ and $p$ are 160 and 1024 bits, respectively. We ran our experiments on macOS High Sierra, 2.3 GHz Intel Core i5, 8GB RAM, and 256GB hard disk. We ran each experiment for 20 times and reported the average.

We used the Flickr dataset in our experiments and using the SALab tool~\footnote{~\href{https://github.com/gaborgulyas/salab}{GitHub: SALab}}, we generated two graphs based on Flickr.  Node and edge similarities are set to 0.5. Graph 1 and Graph 2 has 37377 and 37374 nodes; 1886280 and 1900553 edges respectively. We picked two random nodes, determined their neighbour sets in each graph and ran our experiments using them. In Graph 1, node 1 has 120 neighbours; 48 for node 2 in graph 1; 114 for node 1 in graph 2; and 47 for node 2 in graph 2. Figure~\ref{fig:experiments} shows the total run-time of the common neighbour protocol on the joint graph if we vary the size of neighbours for Node 1 and Node 2 in both graphs.

\subsection{Utility of the Protocol.} 

\begin{figure}[t]
	\centering
	\includegraphics[width=.75\textwidth]{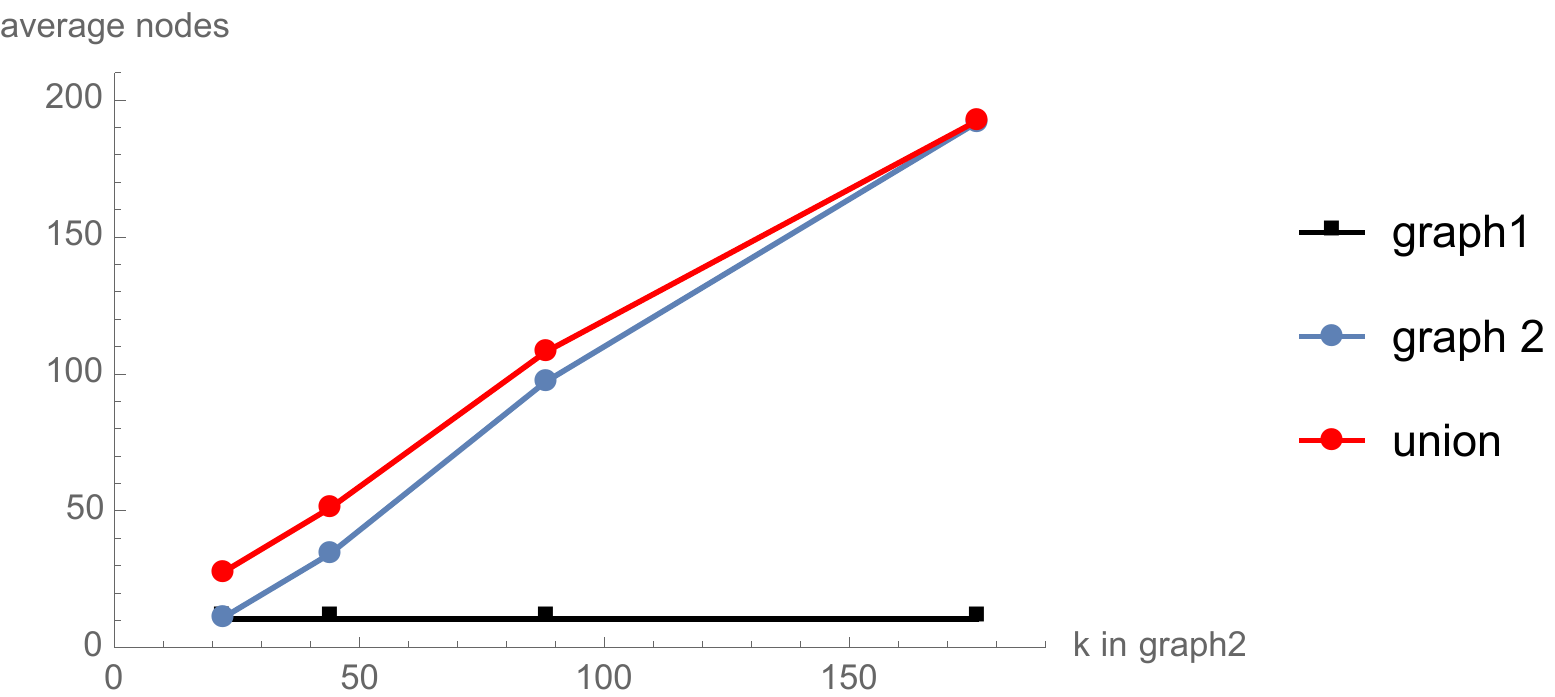}
	\caption{The change in the average number of common neighbours of two pairs according to connectedness of Graph2. $k$ is the number of edges added at each step in Barabasi-Albert distribution. As $k$ increases in Graph 2, Graph 1 benefits more and more from performing the protocol with Graph 2.\label{fig:kvalues}}
\end{figure}

We illustrate the additional common neighbour information gained by a graph when it collaborates with a second graph. In our first experiment, we created two graphs with the same Barabasi-Albert Distribution where 22 edges are added at each step. Both graphs contain 4039 nodes. Average number of common neighbours for each pair is 0.9 in Graph 1. When we consider the merged graph of two networks, average number of common neighbours for each pair increases to 3.3. This shows that distributed link prediction provides significantly more accurate results (compared to local link prediction) and is worth pursuing.

In our second experiment, we created two graphs with the same number of nodes. Both Graph 1 and Graph 2 have 200 nodes. Graph 1 is created with Barabasi-Albert Distribution, where 22 edges are added at each step. Graph 2 is also created in the same setting and, we computed the average number of neighbours of each pair in the union and in Graph 2, with increasing values of k. In this setting, as Graph 2 becomes more connected, it benefits less from the distributed link prediction, as the connectedness of Graph 2 becomes more similar to the union. This is shown in Figure~\ref{fig:kvalues}.

\subsection{Security}

The privacy and integrity of our proposed solution is largely subsumed by the security of the underlying PSI protocol~\cite{de2012fast}. This protocol is shown to be secure under the decisional Diffie-Hellman problem in an appropriate group with semi-honest adversaries. The security proof is in the random oracle model. We make three sequential calls to the protocol. While (universal) composability of the PSI protocol is left by the authors for future work, there is no obvious issue with running the protocol multiple times. For safety, Graph 1 can wait for the first PSI to finish before starting the second one. The PSI protocol leaks (an upper-bound) on the size of each party's graph, and its output is the the cardinality of the intersection. Our protocol, with the three PSI calls, leaks the cardinality of four intermediary values (local2, crossover1, crossover2, and overlap) in computing the common neighbours. We now quantify the impact of this leakage on what Graph 1 can learn about Graph 2 beyond the number of common neighbours.

\paragraph{Leakage of Partial Information.}
\label{sec:leak}

\begin{figure}[t]
	\centering
	\includegraphics[width=.8\textwidth]{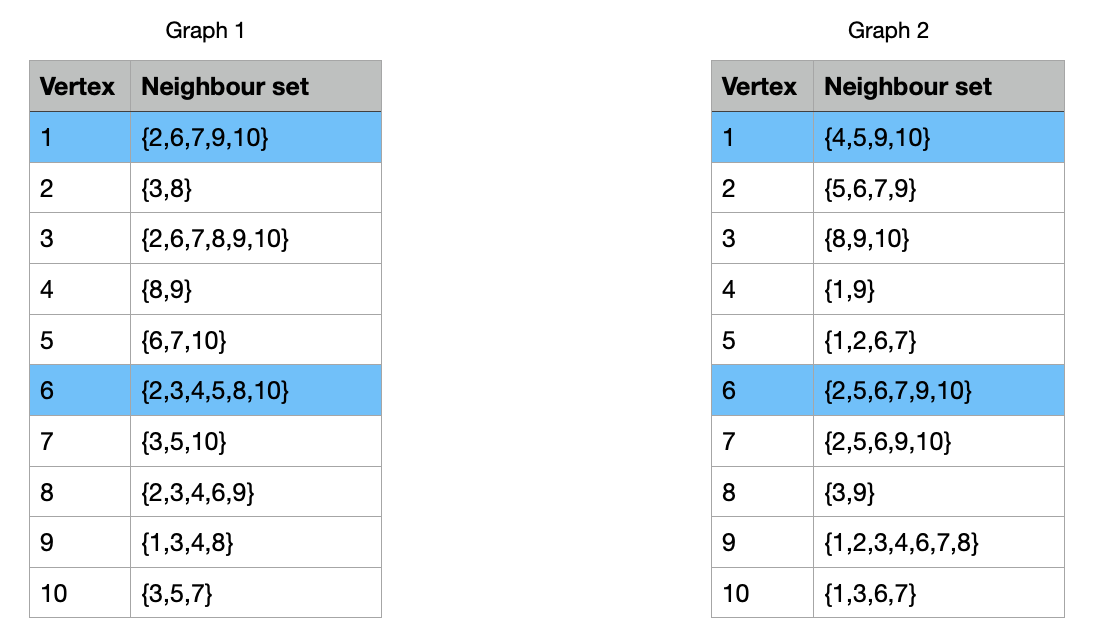}
	\caption{Sample graphs with 10 nodes. We refer to this sample graph in Section~\ref{sec:Backg} to explain how common neighbours are computed among two parties and in Section ~\ref{sec:leak} to quantify the partial information leak.  }
	\label{fig:info}
	\vspace{-10pt}
\end{figure}

\begin{figure}[t]
	\centering
	\includegraphics[width=.75\textwidth]{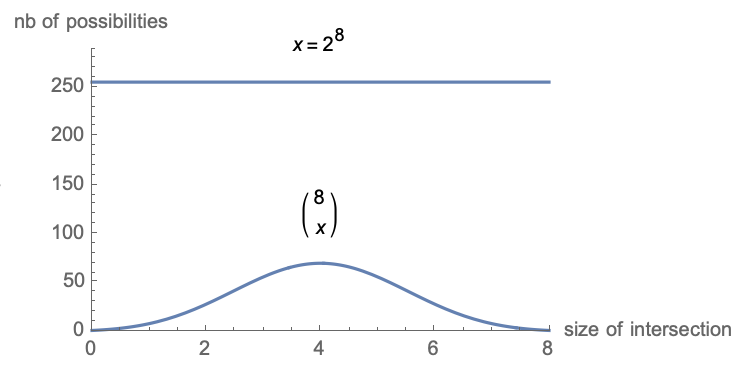}
	\caption{Number of possible combinations of intersection set according to cardinality of intersection.}
	\label{fig:binomial}
	\vspace{-10pt}
\end{figure}

In our setting, there are three different categories of threats to privacy: identity disclosure, link disclosure, and attribute disclosure. As PSI does not leak the nodes in the set intersection, we do not learn about the identities or the attributes related to the nodes. In PSI~\cite{de2012fast}, the cardinality of intersection leaks information about the possible combination of nodes in the intersection set. Graph 1, who learns the size of the intersection set, can compute these combinations (which corresponds to link disclosure). The only time Graph 1 learns the identity of a node (which means that the node is in the set that Graph 2 owns) is when Graph 1 has only one node in its set and the cardinality of the intersection it receives as the result of the PSI protocol is 1. Consider the case where  Node 1 in Graph 1 has only the node 7 in the set, in Figure~\ref{fig:CNexample}. When Graph 1 learns that Crossover2 is 1, it can infer that 7 is connected to Node 6 in Graph 2. Therefore, Graph 1 learns the identity of one of the nodes in the neighbour set of Node 6 in Graph 2 and consequently, the link between 7 and Node 6. 

We refer to the graphs in Figure~\ref{fig:info} in order to illustrate what type of information is leaked during the PSI protocol run between Graph 1 and Graph 2, each having 10 nodes. 
Graph 1 wants to utilize the graph structure of Graph 2 to decide whether to recommend a link between nodes $1$ and $6$. At the beginning of the protocol, Graph 2 sends the size of the common neighbours of node 1 and node 6 (which corresponds to local2  and its size is 3) to Graph 1. Hence, Graph 1 learns that there are $\binom{8}{3}$ possibilities for local2 as opposed to $2^8$ (we choose from 8 nodes as we assume that 1 and 6 are not neighbours in both of the graphs). When we look at the end cases: (i) if the size of intersection is 0, Graph 1 does not learn any extra information (for node 1, there are $2^8$ possibilities with the condition that for each possibility, node 1 and node 6 do not have any nodes in the intersection); and (ii) if the size of the intersection is 8, Graph 1 learns that nodes 1 and 6 are connected to all of the 8 nodes. Figure~\ref{fig:binomial} illustrates the number of possibilities learned by Graph 1 for each size of the intersection. It also shows that even at the worst case, there is still a lot of information that is not learned by Graph 1.

For a graph generated using the Flickr data set that has 37377 nodes, the average number of neighbours of a node is 50. So, for two nodes with average number of neighbours, there are $\binom{37377}{50}$ possibilities for their intersection, which is a very large number.

\section{Discussion}

\subsection{Strengthening the Privacy}
\label{sec:phe}

\begin{figure}[t]
	\centering
	\includegraphics[width=.95\textwidth]{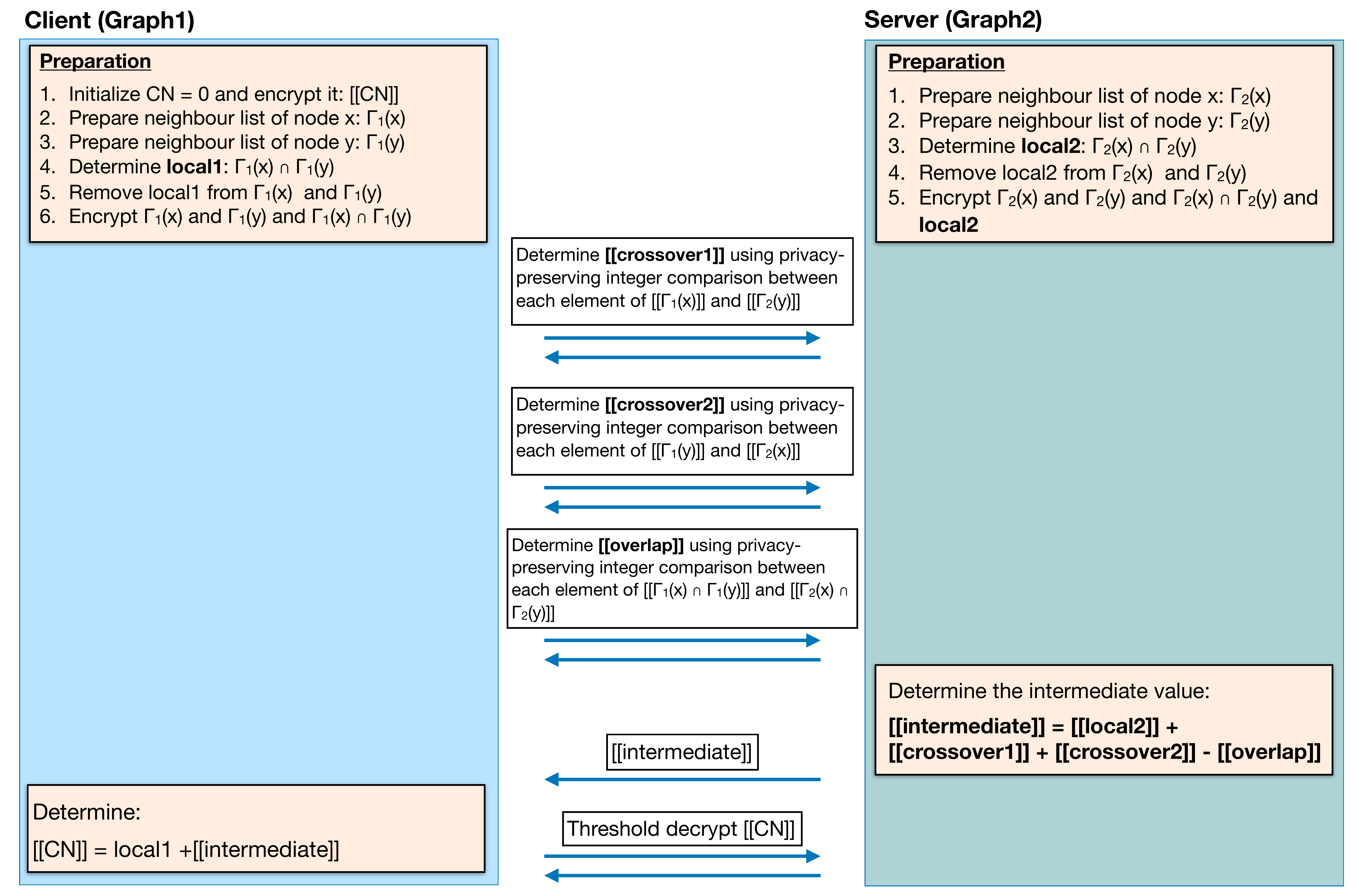}
	\caption{Overview of the solution with stronger privacy. Note that $[[\Gamma_1(x)]]$ means that each element in the neighbour set of node x in Graph 1 is encrypted.\label{fig:privacyenhanced}}
\end{figure}

Here, we discuss a solution based on additively homomorphic encryption (\textit{e.g.,} exponential Elgamal or Paillier) that hides all partial information such that Graph 1 learns only the cardinality. At the beginning of the protocol which is shown in Figure~\ref{fig:privacyenhanced}, Graph 1 determines the neighbour sets of nodes $x$ and $y$, determines local1 and removes local1 from these sets. It encrypts each element in these sets and the cardinality of local1. Graph 2 performs the same steps for local2. In order to determine crossover1, an encrypted  matrix is created: each encrypted element in the neighbour set of node $x$ at Graph 1 is compared against all of the encrypted elements in the neighbour set of node $y$ at Graph 2. The same is done for crossover2 between the neighbour set of $y$ at Graph 1 and the neighbour set of node $x$ at Graph 2. In order to compare two encrypted values, we adapt the protocol proposed in~\cite{health}. The comparison function takes two encrypted values and the output is either the encryption of $0$ if these encrypted values are different or the encryption of $1$ otherwise. The sum over all the elements in the matrix is determined using homomorphic addition both for crossover1 and crossover2. Overlap, which is the intersection of crossover1 and crossover2, is also determined in a similar way using privacy-preserving integer comparison. The final common neighbours cardinality (CN), which is encrypted, is determined as CN= local1 + local2 + crossover1 + crossover2 - overlap. Even though this approach strengthens privacy, it is significantly more costly compared our proposed solution based on PSI. We think for most applications, the efficiency is a larger concern than the partial leakage from our protocol, but it is possible that entities might prefer complete privacy for very sensitive data. 

\subsection{Complexity}
Our protocol's complexity is linear in the sizes of the neighbour set of the nodes. In this paper, we discuss the setting for performing distributed link prediction between two particular nodes in two networks. A network may want to expand this computation for every possible pair in its graph. The complexity depends on the size (which affects the number of possible pairs) and the density of the network (the sizes of neighbour sets affects PSI run-time). PSI runs in linear time complexity. Our protocol is for a specific pair of nodes. The number of pairs in a graph with $n$ nodes is $n^2$, so running our protocol (or any protocol based on PSI) one an entire graph will require running a linear time operation on a quadratic number of nodes: thus, cubic time complexity in the worst-case. Reductions in this complexity (\textit{e.g.,} based on heuristics for prioritizing which nodes to look at) is an interesting future work. 

\section{Concluding Remarks}\label{sec:Conclusion}

For better accuracy, link prediction can be performed on merged graphs belonging to different parties. This leads to privacy concerns as parties do not want to reveal sensitive data related to their network structures. Therefore, in this work, we proposed a PSI-based, privacy preserving distributed link prediction scheme among two graph databases. In our current proposed scheme, Graph 2 learns among which nodes Graph 1 is computing the common neighbours metric. As a future work, a scheme that allows Graph 2 to hide the identities of these nodes from Graph 1 can be proposed. We might also explore if more recent PSI proposals~\cite{karakocc2020linear,chandran2021circuit} produce faster results.

\subsubsection*{Acknowledgements.} We thank the reviewers who helped to improve our paper. J. Clark acknowledges support for this research project from  the National Sciences and Engineering Research Council (NSERC), Raymond Chabot Grant Thornton, and Catallaxy Industrial Research Chair in Blockchain Technologies and NSERC through a Discovery Grant. E. Ayday acknowledges that research reported in this paper was partly supported by the National Science Foundation (NSF) under grant number NSF CCF 2200255 and Cisco Research University Funding grant number 2800379.

 \bibliographystyle{splncs04} 
\bibliography{all}

\begin{thebibliography}{10}
\providecommand{\url}[1]{\texttt{#1}}
\providecommand{\urlprefix}{URL }
\providecommand{\doi}[1]{https://doi.org/#1}

\bibitem{anagreh2021parallel}
Anagreh, M., Laud, P., Vainikko, E.: Parallel privacy-preserving shortest path
  algorithms. Cryptography  \textbf{5}(4), ~27 (2021)

\bibitem{boshrooyeh2017inonymous}
Boshrooyeh, S.T., K{\"u}p{\c{c}}{\"u}, A.: Inonymous: anonymous
  invitation-based system. In: Data Privacy Management, Cryptocurrencies and
  Blockchain Technology, pp. 219--235. Springer (2017)

\bibitem{boshrooyeh2019anonyma}
Boshrooyeh, S.T., K{\"u}p{\c{c}}{\"u}, A., {\"O}zkasap, {\"O}.: Anonyma:
  Anonymous invitation-only registration in malicious adversarial model.
  Cryptology ePrint Archive  (2019)

\bibitem{privpresgraph}
Brickell, J., Shmatikov, V.: Privacy-preserving graph algorithms in the
  semi-honest model. In: Advances in Cryptology-ASIACRYPT 2005. pp. 236--252.
  Springer (2005)

\bibitem{chandran2021circuit}
Chandran, N., Gupta, D., Shah, A.: Circuit-psi with linear complexity via
  relaxed batch opprf. Cryptology ePrint Archive  (2021)

\bibitem{chen2020survey}
Chen, C., Cui, J., Liu, G., Wu, J., Wang, L.: Survey and open problems in
  privacy preserving knowledge graph: Merging, query, representation,
  completion and applications. arXiv preprint arXiv:2011.10180  (2020)

\bibitem{chen2018labeled}
Chen, H., Huang, Z., Laine, K., Rindal, P.: Labeled psi from fully homomorphic
  encryption with malicious security. In: Proceedings of the 2018 ACM SIGSAC
  Conference on Computer and Communications Security. pp. 1223--1237 (2018)

\bibitem{chen2017fast}
Chen, H., Laine, K., Rindal, P.: Fast private set intersection from homomorphic
  encryption. In: Proceedings of the 2017 ACM SIGSAC Conference on Computer and
  Communications Security. pp. 1243--1255 (2017)

\bibitem{de2012fast}
De~Cristofaro, E., Gasti, P., Tsudik, G.: Fast and private computation of
  cardinality of set intersection and union. In: International Conference on
  Cryptology and Network Security. pp. 218--231. Springer (2012)

\bibitem{dong2012link}
Dong, Y., Tang, J., Wu, S., Tian, J., Chawla, N.V., Rao, J., Cao, H.: Link
  prediction and recommendation across heterogeneous social networks. In: Data
  Mining (ICDM), 2012 IEEE 12th International Conference on. pp. 181--190. IEEE
  (2012)

\bibitem{dong2015coupledlp}
Dong, Y., Zhang, J., Tang, J., Chawla, N.V., Wang, B.: Coupledlp: Link
  prediction in coupled networks. In: Proceedings of the 21th ACM SIGKDD
  International Conference on Knowledge Discovery and Data Mining. pp.
  199--208. ACM (2015)

\bibitem{duong2020catalic}
Duong, T., Phan, D.H., Trieu, N.: Catalic: Delegated psi cardinality with
  applications to contact tracing. In: International Conference on the Theory
  and Application of Cryptology and Information Security. pp. 870--899.
  Springer (2020)

\bibitem{freedman2004efficient}
Freedman, M.J., Nissim, K., Pinkas, B.: Efficient private matching and set
  intersection. In: International conference on the theory and applications of
  cryptographic techniques. pp. 1--19. Springer (2004)

\bibitem{he2009efficient}
He, X., Vaidya, J., Shafiq, B., Adam, N., Terzi, E., Grandison, T.: Efficient
  privacy-preserving link discovery. In: Pacific-Asia Conference on Knowledge
  Discovery and Data Mining. pp. 16--27. Springer (2009)

\bibitem{health}
Hubaux, J.P., Fellay, J., Ayday, E., Laren, M., Raisaro, J., Jack, P., et~al.:
  Privacy-preserving computation of disease risk by using genomic, clinical,
  and environmental data. In: Proceedings of USENIX Security Workshop on Health
  Information Technologies (HealthTech” 13), number EPFL-CONF-187118 (2013)

\bibitem{karakocc2020linear}
Karako{\c{c}}, F., K{\"u}p{\c{c}}{\"u}, A.: Linear complexity private set
  intersection for secure two-party protocols. In: International Conference on
  Cryptology and Network Security. pp. 409--429. Springer (2020)

\bibitem{leicht2006vertex}
Leicht, E.A., Holme, P., Newman, M.E.: Vertex similarity in networks. Physical
  Review E  \textbf{73}(2),  026120 (2006)

\bibitem{linkepred}
Liben-Nowell, D., Kleinberg, J.: The link-prediction problem for social
  networks. Journal of the American society for information science and
  technology  \textbf{58}(7),  1019--1031 (2007)

\bibitem{liu2010link}
Liu, W., L{\"u}, L.: Link prediction based on local random walk. EPL
  (Europhysics Letters)  \textbf{89}(5),  58007 (2010)

\bibitem{pinkas2019efficient}
Pinkas, B., Schneider, T., Tkachenko, O., Yanai, A.: Efficient circuit-based
  psi with linear communication. In: Annual International Conference on the
  Theory and Applications of Cryptographic Techniques. pp. 122--153. Springer
  (2019)

\bibitem{tangtransfer}
TANG, J., LOU, T., KLEINBERG, J., WU, S.: Transfer learning to infer social
  ties across heterogeneous networks

\bibitem{surveyofprivacy}
Wu, X., Ying, X., Liu, K., Chen, L.: A survey of privacy-preservation of graphs
  and social networks. In: Managing and mining graph data. pp. 421--453.
  Springer (2010)

\bibitem{xu2019privacy}
Xu, Z., Zhou, F., Li, Y., Xu, J., Wang, Q.: Privacy-preserving subgraph
  matching protocol for two parties. International Journal of Foundations of
  Computer Science  \textbf{30}(04),  571--588 (2019)

\bibitem{yu2008gaussian}
Yu, K., Chu, W.: Gaussian process models for link analysis and transfer
  learning. In: Advances in Neural Information Processing Systems. pp.
  1657--1664 (2008)

\bibitem{zhang2014meta}
Zhang, J., Yu, P.S., Zhou, Z.H.: Meta-path based multi-network collective link
  prediction. In: Proceedings of the 20th ACM SIGKDD international conference
  on Knowledge discovery and data mining. pp. 1286--1295. ACM (2014)

\end{thebibliography}

 \appendix

 \section{Details of Comparison of Cryptographic Primitives}
 \label{app:crypto}

 Recall Table~\ref{Table:crypt}. In this appendix, we provide further detail. De Cristofaro \cite{de2012fast} introduced a method to calculate PSI-CA in which the client or the server learns no information about each others' private input. Moreover, the two instances of the protocol on the same inputs are unlinkable. Beyond revealing the cardinality to the client, the upper bound of each set is the only information leaked to the parties. The protocol is secure against semi-honest adversaries who follow the protocol's instructions. The complexity remains linear in the size of the two sets.

 In some scenarios, where the computation power of the parties is limited, they delegate the computation to an untrusted cloud server. Duong \emph{et al.} \cite{duong2020catalic} suggested a delegated PSI-CA based on oblivious distributed key PRF (Odk-PRF). The protocol has two phases. In the first phase, the party who wants to receive the cardinality result (in our scenario, Graph 1) distributes their input's secret shares to the cloud server. The Odk-PRF protocol is deployed between the cloud server and the sender party (who is Graph 2). As a result, the cloud server obtains secret shares of the PRF output for each share of Graph 1's input, and Graph 2 learns the combined PRF key. In the second phase, Graph 2 generates a set of key-value pairs using their own inputs and Graph 1's input shares. The cloud server can obliviously search graph 2's key-value pairs and obtain correct values known to Graph 1. The server adds some fake values, permutes, and sends them to graph 1, which can check the number of items in the intersection (PSI-CA) by
 counting the "real" values obtained from the cloud server. The protocol is secure against semi-honest adversaries who can corrupt parties. 
 Finally, the cardinality result is revealed to Graph 1, and the secret shares of Graph 1 are revealed to the cloud server. Moreover, as a result of running the Odk-PRF protocol with Graph 2, The cloud server views a subset of the PRF value. Graph 2 has PRF key shares. Although these values reveal no information about the parties' secret/key shares, the setting requires non-colluding parties. The computation and communication complexity of the protocol is linear in the size of the smaller set. 

 Chen \emph{et al.} \cite{chen2017fast}, developed a PSI protocol based on fully
 homomorphic encryption. Graph 1 holds a set $X$ with the size of $N_x$, and Graph 2 has set $Y$ with the size $N_y$. These sets consist of $\sigma$-bit strings. Graph 1 encrypts each element of its set and sends them to Graph 2 who homomorphically evaluates the product of differences of these values with all of the Graph 2’s items. Graph 2 randomizes the results by multiplying the items with a uniformly random non-zero plaintext, and sends the result back to Graph 1. The latter decrypts the results to zero when there exists a common item in Graph 2’s set. Otherwise, the result returns a uniformly random non-zero plaintext.  The security holds in the semi-honest model.
 This protocol leaks no information about the sets to the parties. Apart from revealing the intersection result to Graph 1, the set sizes, the (common) bit-length $\sigma$ are publicly known values. 
 The communication complexity of the protocol is linear in the size of the smaller set, and logarithmic in
 the larger set.

 Pinkas \emph{et al.} \cite{pinkas2019efficient} developed a solution for PSI based on oblivious
 programmable pseudo-random functions (OPPRF). There protocol comprises of a circuit that its input wires are associated with the hash of input values in Graph 1 and Graph 2's sets. If the items in both sets are equal the circuit's output wire returns $1$, and $0$, otherwise. The circuit computes the number of $1$'s and reveals it.
 This solution \cite{pinkas2019efficient} reveals the set intersection result and bit-length of the items to both parties. 
 The security is modeled for malicious parties. Moreover, the complexity is linear.

 Chen \emph{et al.} \cite{chen2018labeled} developed a labeled PSI-based solution based on OPRF and homomorphic encryption. In this setting, Graph 2 holds a label $l_i$ for each item in its set. Graph 1 will also learn the labels corresponding to the items in the intersection. The parties use OPRF in a pre-processing phase to hash their values. Then, they deploy FHE to obtain the final result.
 The complexity of the scheme remains logarithmic in the size of larger set and linear in the size of smaller set.
 While the intersection result and a label are delivered to Graph 1, the number of items in each set after hashing is leaked to public. The security of the scheme is proven for malicious adversaries.

\end{document}